\documentclass[prd,reprint]{revtex4}
\usepackage{amsmath}
\usepackage{amsbsy}
\usepackage{hyperref}
\usepackage[english]{babel}
\usepackage{natbib}

\def\cu{{\cal U}}
\def\cp{{\cal P}}
\def\cm{{\cal M}}
\def\cs{{\cal S}}
\def\fbh{F_{\mbox{\tiny{BH}}}}
\begin{document}

\title{Moving Schwarzschild Black Hole and Modified Dispersion Relations}

\author{Cristian Barrera Hinojosa$^a$\footnote{cbarrera.hinojosa@gmail.com}, Justo L\'opez-Sarri\'on$^{a,b}$\footnote{jujlopezsa@unal.edu.co}}

\affiliation{$^a$Departamento  de  F\'{\i}sica,  Universidad  de  Santiago  de
  Chile, Casilla 307, Santiago, Chile\\
$^b$Departamento  de  F\'{\i}sica, Universidad  Nacional de Colombia, 111321,  Bogot\'{a}, Colombia}

\begin{abstract}

We study the thermodynamics of a moving Schwarzschild black hole, identifying the temperature and entropy in a relativistic scenario. Furthermore, we set arguments in a framework relating invariant geometrical quantities under global spacetime transformations and the dispersion relation of the system. We then extended these arguments in order to consider more general dispersion relations, and identify criteria to rule them out.

\end{abstract}
\maketitle

\section{Introduction}

The search  for an unified  picture of gravity and quantum principles
has deserved a lot of  efforts  and the approaches to this
problem have taken  different roads  and  explored a  wide class  of
scenarios. Theoretical proposals like String Theory, Loop Quantum Gravity, or Causal Sets are examples of that, and from there, one tries  to find  clues of some  characteristic feature  which might survive at  energy scales today  attainable in observed  phenomena, or some  tiny effects  together with  a sort  of amplification  mechanism which could offer evidence. In this regard,  the modification of the dispersion  relation of particles has received much attention in last years \cite{Mattingly, Piran, Magueijo} as a possible signal of quantum  gravity effects  in  the  context of  the  so called  quantum gravity  phenomenology, namely,  the  search of  observable relics of quantum gravity \cite{AmelinoLIVREV}. Among the  proposals sharing this  disctintive feature we  can mention some  limits  of  Loop  Quantum  Gravity \cite{Alfaro},  Double  Special Relativity \cite{Amelino2000DSR, Amelino2002DSR},  String  theory \cite{String-MDR}, etc. In  this context,  the  dispersion relation  that  a macroscopic  body satisfies --  taken as  a body whose  constituents satisfy  a modified
dispersion relation  --  might be a  source of difficulties when one  adopt this point of view, as noted already  in early stages of these proposals \cite{Soccerball2002}. 

By  other hand,  black holes  are probably  one of  the most  suitable objects in nature  to test -- at least theoretically  -- some features of  quantum  gravity.  For  example, logarithmic  corrections  to  the area-entropy  law  seems to  be  a  common characteristic  shared  by different proposals of quantum gravity \citep{Kaul, Frolov,Carlip}.

Under these considerations it seems natural to analyze if some imprints of quantum gravity effects could be codified in the dispersion relation that a black hole satisfies \citep{Amelino}. In this work we will try to gain some insight in this matter, studying the kinematic consequences of a generalization of symmetries in the movement of a massive object as a black hole. 

The layout of this paper is the following:  in the present section we motivate the problem and the questions that we would like to answer. In section II we study the metric of a moving Schwarzschild black hole, or the Sexl-Aichelburg metric, and calculate its temperature. Furthermore, in that section we will generalize the geometrical thermodynamic arguments to this moving black hole, identifying relevant parameters which will be useful later. In Section III we extend the formalism carried out in the previous section to take into account dispersion relations beyond the Lorentzian one. We will find certain consistency relations that those modified relations have to fulfill. In section IV we briefly discuss some interesting examples found in the literature. Finally, we conclude in section V with final remarks, observations and outlook.

\section{Moving Schwarzschild Black Hole}
In this section we will analyze a Schwarzschild black hole moving with respect to an observer, which will be carried out by means of a Lorentz boost to the stationary solution. We calculate the geometrical temperature and analyze the structure of its partition function.

\subsection{Temperature of the moving black hole}

Let us consider a black hole described by the Schwarzschild metric in coordinates $(t',r',\theta',\varphi')$
\begin{eqnarray}
\label{schw}
ds^2 &=&- \left(1- \frac{2M}{r'}\right)dt'^2 + \frac{dr'^2}{1-\frac{2M}{r'}} + r'^2d\Omega'^2,
\end{eqnarray}
 where $M$ is the mass and we are using prime variables for the stationary observer. In order to {\it put the black hole in motion}, we follow the approach of Aichelburg and Sexl \cite{achsex}. There, 
authors consider an observer moving uniformly relative to the mass $M$ who sees a metric which is the previous one
deformed by a Lorentz transformation. That is, if $v$ is the velocity of the black hole in the $z$ direction and $\gamma=(1-v^2)^{-1/2}$ the relativistic factor, we must take in (\ref{schw}) $t' = \gamma(t-vz)$ and $r'=\sqrt{x^2+y^2+\gamma^2(z-vt)^2}$. 

In order to study the thermodynamics of this object we will follow the path integral approach of Euclidean path integral \cite{hawking} along with an identification of spacetime points according to the system in movement. The action one considers is the regularized gravitational action
\begin{equation}
\label{regact}
I=\frac{1}{16\pi}\int_V\sqrt{-g}R\,d^4x +\frac{1}{8\pi} \int_{\partial V} [K]\sqrt{-h}\,d^3x,
\end{equation}
where $R$ is the scalar curvature, $g$ the metric determinant, $[K]$ the trace of the extrinsic curvature regularized with respect to the flat spacetime and $h$ the determinant of the induced metric on the boundary $\partial {V}$ of the volume $V$. We will take $\partial {V}$ as the Lorentz transformed of the stationary boundary for the Schwarzschild solution, i.e. $r'=r_0'>2M$, rendering $r_0' \to \infty$.

The value of the  Euclidean action can be obtained straightforwardly by noticing that both integrands in (\ref{regact}) are scalars under coordinate transformations,
and, as pointed out before, we are integrating over the boosted Schwarzschild boundary. Then, the value of the action is the same as computed in stationary coordinates with the Schwarzschild boundary, i.e.
\begin{equation}
\label{actionvalue}
I=4\pi iM^2.
\end{equation}

By the other hand, in spite that the value of the Euclidean action does not change, the temperature does. In fact,  the partition function 
in the path-integral approach must be integrated over periodic paths. Since the system is moving in the $z$ direction with velocity $v$, the periodicity condition implies the identification of points in spacetime as $(t,x,y,z)\sim(t+i\beta,x,y,z+iv\beta)$. If we make the following coordinate transformation in (\ref{schw}),
\begin{eqnarray}
\frac{y-i\xi}{y+i\xi}&=&e^{t'/2M},\label{K1}\\
y^2+\xi^2&=&\left(\frac{r'}{2M}-1\right)e^{r'/2M},
\end{eqnarray}

the resulting Kruskal metric is free of conical singularities at $r'=2M$, for imaginary time $t'=-i\tau'$, if the periodicity condition are consistent with the previous transformation, namely $\Delta \tau'=8\pi M$ and $\Delta r'=0$. This implies that  
\begin{equation}\label{deltatau'}
\Delta \tau'=i\gamma(\Delta t-v\Delta z)=\gamma\beta(1-v^2)=8\pi M,
\end{equation}

where $t=-i\tau$. Hence, from (\ref{deltatau'}) we deduce that
\begin{equation}\label{chap2:beta}
\beta=8\pi M \gamma,
\end{equation}

which is consistent for the case of Lorentz invariant systems \cite{Tolman}. Furthermore, $\Delta z'=\gamma (\Delta z-v\Delta t)=0$, and then $\Delta r'=0$ as required.

Alternatively, one can use the Unruh Effect \cite{Hawkingevap,Unruhevap} in order to find the temperature of the system in movement as follows. Under the transformations $\tilde{\tau}=t'/(4M)$, $\rho^2=8M(r'-2M)$ and $\tilde{x}=(2M\theta',2M\phi')$, the metric (\ref{schw}) can be approximated in the vicinity of any point ($t_0,x_0,y_0,z_0$) of the event horizon $r'=2M$ as the Rindler metric
\begin{equation}
ds^2\approx-\rho^2d\tilde{\tau}^2+d\rho^2+d\tilde{x}^2,
\end{equation}

which describes an accelerated frame, with acceleration $(4M)^{-1}$ respecting the primed coordinates. If we focus on a moving point on the horizon (with respect to the unprimed coordinates) we have $\tilde{\tau}=\frac{\gamma}{4M}(t-vz)$. Then, in an interval of time $\Delta t$, we have that $\Delta z=v\Delta t$, i.e. $\Delta\tilde{\tau}=\frac{\gamma}{4M}(1-v^2)\Delta t=\frac{\Delta t}{4M\gamma}$. Hence, the surface gravity measured on the unprimed system is $\kappa=(4M\gamma)^{-1}$, and then, by means of the Unruh Effect, the temperature associated to the Hawking Radiation of the moving black hole is $T=\beta^{-1}=\kappa(2\pi)^{-1}=(8\pi M \gamma)^{-1}$, as given by (\ref{chap2:beta}). Notice that a global boost on the stationary system  does not correspond to a local boost in the proximity of the event horizon, but to a change in the surface gravity.

\subsection{Thermodynamics and Lorentz Symmetry}

We now follow thermodynamical arguments to discuss the moving black hole problem. The system is described by an observer ${\cal O}$ such that  it has conserved  linear  momentum ${\cal P } $ and Hamiltonian $H$ (for simplicity,  we consider the case where the relative velocity is aligned with one axis of the reference system of  the distant observer). In this scenario, we can write the Massieu function as
  
\begin{equation}
\label{FLI}
F(\beta,\eta)=-\ln Z=-\ln\left[\mbox{Tr}\left(e^{-\beta H + \eta P}\right)\right],
\end{equation}

where $\beta$ and $\eta$ are chemical potentials (Lagrange  multipliers) associated  to  the  conservation of charges. Notice that $F$ will coincide numerically with the path integral formalism with imaginary time of quantum mechanics. Its Legendre transformed function corresponds to the entropy $S({\cu}, {\cp})$, given by

\begin{equation}
\label{entropyLI}
S({\cal U}, {\cal P}) = - F(\beta,\eta) + \beta\,{\cu} -\eta\,{\cp},
\end{equation}

with

\begin{equation}
\label{UQ}
\cu =\left(\frac{\partial F}{\partial \beta}\right)_{\eta}, ~~~~~~~~~
\cp = -\left(\frac{\partial F}{\partial \eta}\right)_{\beta}.
\end{equation}

We are interested in the connection that can be established between the entropy of the system and the dispersion relation it satisfies. Given that dispersion relations are associated to spacetime symmetries of the system, we will start our analysis by considering the case of Lorentz invariance and we will observe how, by imposing this, it is possible to connect the functional dependence of $S$ and $F$ to certain invariants quantities built from the intensive variables $\beta,\eta$. If the thermodynamical description done by an observer ${\cal O}'$ -- at rest with the system -- is related with the description
done by ${\cal  O}$ through the Lorentz  transformations on quantities
$H,P,\beta,\eta$, then  $F$ must depend on a particular combination 
of the variables $(\beta,\eta)$ which remains invariant. Indeed, let us consider an unitary operator $U(\epsilon)$ which represents a Lorentz transformation of relative rapidity $\epsilon$, then
\begin{eqnarray}\label{Ftransformada}
F(\beta,\eta) &=& -\ln\left[\mbox{Tr}\left (U^\dag e^{-\beta H + \eta P}U\right)\right] 
 \nonumber
 \\
&=&-\ln\left[\mbox{Tr}\left (e^{-\beta H' + \eta P'}\right)\right] 
\nonumber
\\
&=&
-\ln\left[\mbox{Tr}\left (e^{-\beta' H + \eta' P}\right)\right] \equiv F(\beta',\eta').
\nonumber
\end{eqnarray}

where we have used fact that $H$ and $P$ transform linearly under the Lorentz group, i.e.
\begin{eqnarray}
H' = U^\dag H U=H\cosh\epsilon-P\sinh\epsilon\\
P' = U^\dag P U=P\cosh\epsilon-H\sinh\epsilon,
\end{eqnarray}

and then
\begin{eqnarray}
\beta'=\beta\cosh\epsilon+\eta\sinh\epsilon\\
\eta'=\eta\cosh\epsilon+\beta\sinh\epsilon.
\end{eqnarray}

In the reference system ${\cal O}'$ we have ${\cal P }' =0$, and then $\tanh\epsilon=v$ is the relative velocity between ${\cal O}$ and this observer, i.e. $\cosh\epsilon=\gamma$, and $\sinh\epsilon=\gamma v$ and therefore, in the present case, the function $F$ must depend on $(\beta,\eta)$
through the combination $\beta^2 -\eta^2$.
 Consider now the particular
function $\mu(\beta,\eta)$ defined through
\begin{equation}
\label{mu}
\mu^2=\beta^2 -\eta^2,
\end{equation}
which invariant by construction. Notice that this quantity $\mu$ corresponds to the temperature of the system at rest ($\eta=0$, $\beta=\beta_0$). For the black hole under consideration, this quantity has a topological origin, in virtue of the periodicity condition, and then it is appropriate to relate it to the Schwarzschild radius $R_s$, concretely

\begin{equation}\label{muyR}
\mu=\beta_0=4\pi R_s.
\end{equation}

This is consistent with the fact that since the area of the event horizon $A$ is a marginally trapped surface, it will not change under a boost of the Schwarzschild metric (\ref{schw}), and from (\ref{muyR}) one finds the relation 
\begin{equation}
\label{muarea}
\mu^2 = 4\pi A.
\end{equation}

On the other hand, since the internal energy ${\cal U}$ and momentum ${\cal P}$ are related to $F$ through (\ref{UQ}) one obtains (for the invariant (\ref{mu})) the following relations 
\begin{equation}\label{cucp}
\cu = \dot{F}(\mu)\frac{\beta}{\mu}\,,~~~~~~~~~
\cp = -\dot{F}(\mu)\frac{\eta}{\mu}\,
\end{equation}
with $\dot{F} \equiv dF/d\mu$. Therefore, it is straightforward to obtain the relativistic dispersion relation

\begin{equation}
\cu^2-\cp^2=\cm^2,
\end{equation}

where we have denoted $\cm=\dot{F}$, which is the inertial mass of the black hole. Notice that in the previous argument, in addition to showing that the internal energy and momentum of the black hole satisfy the relativistic dispersion relation, we have found a nontrivial relation between the inertial mass of the black hole and the Schwarzschild radius through the partition function of the system. On the other hand, using equations (\ref{cucp}) and (\ref{entropyLI}) we obtain

\begin{equation}\label{Smu}
S=-F(\mu)+\mu\dot{F},
\end{equation}

which expresses the entropy as the Legendre transform of $F$ with respect to the invariant $\mu$. It is worth noting that entropy here appears as a relativistic invariant, and its invariance under Lorentz transformations is consistent  with the approach to thermodynamics of moving objects by Tolman, for example \cite{Tolman}. This fact that appears trivial here will have interesting consequences that will be discussed later. Moreover, expressing everything in terms of $\cm$ we obtain the relation $\mu=S'(\cm)$, where $S'$ is the derivative of the entropy with respect to $\cm$. We can read the velocity of the system as

\begin{equation}
v\equiv \left(\frac{\partial U}{\partial P}\right)_S=\frac{\eta}{\beta}
\end{equation}

this is to say $\eta=\beta v$, and using (\ref{mu}) we obtain $\beta=\mu\gamma$. Let us now identify the partition function of this system with the euclidean action in the path integral formalism in the semiclassical approximation. The value of the action (\ref{actionvalue}) together with (\ref{muyR}), and $R_s=2M$ imply that the Massieu function describing the black hole is

\begin{equation}
\label{bhf}
\fbh=\frac{\mu^2}{16\pi},
\end{equation}

and then $\cm=\dot{F}=M$, namely, in the semiclassical approximation the inertial mass coincides with the mass parameter in the Schwarzschild solution. As a consequence, $\cm^2=(R_s/2)^2=A/(16\pi)$, and then we obtain the Smarr formula \cite{smarr,Bardeen1973} for the moving black hole

\begin{equation}
\label{smarlopez}
\cu^2 = \frac{A}{16\pi} +\cp^2.
\end{equation}

Finally, the entropy can be calculated with (\ref{Smu}), and it corresponds to $S=A/4$, and provided that $\mu=8\pi M$ the inverse of the temperature is then $\beta=8\pi M\gamma$, as we proved by other arguments in the previous subsection.

It is natural to expect that both the Bekenstein-Hawking entropy and the Smarr formula will acquire corrections as one takes into account terms beyond the semiclassical approximation. In fact, several approaches to quantum gravity show corrections to the entropy-area law \cite{Medved, Kaul, Frolov, Carlip, Chatterjee}, which can be used to calculate $F(\mu)$ starting from (\ref{muarea}) and (\ref{Smu}) as a differential equation. Given $S=S(A)$, the solution is

\begin{equation}
\label{solf}
F(\mu) = a\,\mu  + \frac{\mu}{\sqrt{16\pi}}\int_{1}^{\mu^2/4\pi}\frac{\cs(A')}{A'^{3/2}}dA',
\end{equation}
from which we also get
\begin{equation}
\label{entrodisp}
\cm(\mu)=a + \frac{\cs(\mu)}{\mu} + \frac{1}{\sqrt{16\pi}}\int_{1}^{\mu^2/4\pi}\frac{\cs(A')}{A'^{3/2}}dA'.
\end{equation}

In particular, certain quantum gravity models, such as Loop Quantum Gravity\cite{Ashtekar, Rovelli} and String theory\cite{Strominger} coincide in that the corrections to the area-entropy law have the form
\begin{equation}
\label{log}
S=\frac{A}{4} +\rho\ln(A),
\end{equation}

where $\rho$ is a model dependent dimensionless quantity of order unity. Using equations (\ref{solf}) and (\ref{entrodisp}) for such case we obtain

\begin{equation}
\label{floga}
F(\mu) = \frac{\mu ^2}{16 \pi }  -2 \rho\left(1 +\log\left(\frac{\mu}{4\pi}\right)\right)+a(\rho)\mu,
\end{equation}

and

\begin{equation}
\label{otrom}
\cm=\ \frac{\mu}{8\pi} - \frac{2\rho}{\mu}+a(\rho).
\end{equation}

Here we have made explicit that the integration constant $a$ may depend on the parameter $\rho$ and, concretely, it is direct to fix its value  when $\rho=0$ in order to reproduce the semiclassical result. Considering the identification of $\mu$ in terms of the Schwarzchild radius, equation (\ref{otrom}) provides us two solutions given by

\begin{equation}
r_\pm= (\cm-a)\left(1\pm\sqrt{1+\frac{\rho}{\pi (\cm-a)^2}}\right),
\end{equation}

where $r_+$ is continuously connected to the semiclassical Schwarzschild radius, and $r_-$ to the singularity at the origin (given that $\rho<0$ independently of the model \cite{Kaul}). This can be observed by expanding both solutions in $\rho/M$ powers
\begin{eqnarray}\label{rmasmenos}
r_+&=&2(\cm-a) +\frac{\rho}{2\pi(\cm-a)}+\mathcal{O}\left(\frac{\rho}{\cm^3}\right),\\
r_-&=&-\frac{\rho}{2\pi(\cm-a)}+\mathcal{O}\left(\frac{\rho}{\cm^3}\right).
\end{eqnarray}

These results resembles the ones obtained in \cite{Reuter1}, where authors start from an effective metric whose Newton constant $G_N$ acquires quantum corrections depending on $r$. In order to make contact with this idea, let us assume that the inertial mass of the black hole has corrections with respect to the semiclassical approximation, i.e.

\begin{equation}
\cm=M+\rho\delta_\cm.
\end{equation}

Writing equation (\ref{rmasmenos}) in terms of $M$ and collecting  powers of $\rho/M$ we can compare with the results obtained in \cite{Reuter1, Donoghue}, identifying $a=\rho\delta_\cm$, and $\rho=-118/15$. It is worth noting that, even if there is some discrepancy with the precise value obtained by other approaches,  both sign and order of magnitude of the model dependent parameter $\rho$ are consistent.

\section{Generalization to Modified Dispersion Relations}

Consider now the possibility that Lorentz symmetry is modified or even broken, as the scenarios we mentioned in the first section, this is to say, we will assume the existence of a deformed symmetry underlying the theory. We ask how previous arguments will change in such case, expecting to be able to identify the invariant $\mu$ in terms of $(\beta,\eta)$ with a more general functional dependence.

A similar argument to equation (\ref{Ftransformada}) implies that the partition function will be invariant under transformations of the form $\beta'=f_\epsilon(\beta,\eta)$, and $\eta'=g_\epsilon(\beta,\eta)$ where $\epsilon$ is any symmetry parameter satisfying that $\beta=f_0(\beta,\eta)$ and $\eta=g_0(\beta,\eta)$. Particularly, if we call $\mu=\beta$ when $\eta=0$, and assume that the transformation connects the points in the ($\beta$,$\eta$) plane in a qualitatively similar form as the Lorentz Transformations, then we can always write down 
\begin{eqnarray}
\beta=f_\epsilon(\mu,0)\\
\eta=g_\epsilon(\mu,0), 
\end{eqnarray}

such that $\mu=f_0(\mu,0)$ and $0=g_0(\mu,0)$. From here we should obtain expressions such that 
\begin{eqnarray}
\mu=\mu(\beta,\eta)\label{mugen}\\
\epsilon=\epsilon(\beta,\eta),
\end{eqnarray}

where $\mu(\beta,0)=\beta$ and $\epsilon(\beta,0)=0$. Finally, we can replace the somewhat ambiguous parameter $\epsilon$ with the compatible parameter $v\equiv\eta/\beta$.

Now, let us suppose that $F(\mu)$ is known.  In such case, we can write
$$
\cu = \dot{F}\frac{\partial\mu}{\partial\beta},~~~~~~~~~
\cp = -\dot{F}\frac{\partial\mu}{\partial\eta},
$$
and in principle, variables $(\beta,\eta)$ can be obtained as function of 
$(\cu,\cp)$.  We define the dispersion relation as 
\begin{equation}
\label{Mgen}
m(\cu,\cp)\equiv\mu(\beta(\cu,\cp),\eta(\cu,\cp)).
\end{equation}

The entropy, by other hand, turns out to be
\begin{equation}
\label{sgen}
S(\cu,\cp) = -F(\mu) +\left(\beta\,\frac{\partial\mu}{\partial\beta} + \eta\,\frac{\partial\mu}{\partial\eta}\right)\dot{F}.
\end{equation}

Notice the following important point: if the expression in parenthesis in equation (\ref{sgen}) depends upon ($\beta$,$\eta$) in an arbitrary way, and not through the invariant $\mu$, then the entropy would no longer be a movement invariant, and that can carry us to some inconsistencies as, for example, the number of accessible states being dependent upon the movement state of the system. Then, we will take as an hypothesis the invariance of the entropy, and consequently we must impose the term in parenthesis to be, in principle, an arbitrary function of $\mu$. However, let us remember that $\mu=\beta$ at rest, and in such case $S=-F+\beta\cu=-F+\mu\dot{F}$. For compatibility of the definitions of entropy in movement and at rest, it is only possible to impose that

\begin{equation}\label{L}
\beta\,\frac{\partial\mu}{\partial\beta} + \eta\,\frac{\partial\mu}{\partial\eta}=\mu.
\end{equation}

This last equation is a particular case of the known Clairaut equation, whose general solution is given by Euler's homogeneous function theorem of degree 1. Then, from equation (\ref{mugen}) and the discussion in the paragraph following it, we can deduce that 
\begin{equation}\label{Gamma}
\mu=\beta\mu(1,v)=\frac{\beta}{\Gamma(v)},
\end{equation}

where we have defined $\Gamma(v)=\mu^{-1}(1,v)$, which satisfies that $\Gamma(0)=1$, and corresponds to a generalization of the relativistic $\gamma$ factor.
In such case, the entropy depends on the dispersion relation and the following relations hold
\begin{eqnarray}
\label{Mmu}
\cm(\mu) &=&\frac{ dF}{d\mu},
\\
\label{muM}
\mu({\scriptstyle \cm})&=& \frac{d\cs}{d{\scriptstyle\cm}}.
\end{eqnarray}

Notice that this last result implies that the discussion at the end of the previous section remains valid even for dispersion relations beyond the relativistic one.

Now, we may ask the inverse problem, namely, knowing the entropy of the system as a function of its internal energy and momentum, compatible with a certain given dispersion relation 

\begin{equation}\label{mgen}
\cm=\cm(\cu,\cp),
\end{equation}

that remains invariant at constant entropy, and where $\cm$ reduces to the rest energy, i.e. $\cu=\cm(\cu,0)$. Then, using the inverse Legendre transform for $S$, and imposing that $F$ depends on an invariant $\mu=S'$,  we have that

\begin{equation}
F=-S+\left(\cu\frac{\partial \cm}{\partial\cu}+\cp\frac{\partial\cm}{\partial\cp}\right)S'=-S+\cm S'.
\end{equation}

This condition implies that

\begin{equation}\label{clairautm}
\cu\frac{\partial \cm}{\partial\cu}+\cp\frac{\partial\cm}{\partial\cp}=\cm,
\end{equation}

in other words, $\cm$ has to be a homogeneous function of degree 1 with respect to $\cu$ and $\cp$. This, again, imposes severe constraints on the form that could have the dispersion relation of a thermodynamical system. 

Given a dispersion relation $\cm(\cu,\cp)$ satisfying (\ref{mgen}) y (\ref{clairautm}) we can write 
\begin{eqnarray}
\beta&=&\mu\frac{\partial \cm}{\partial \cu}\\
\eta&=&-\mu\frac{\partial \cm}{\partial \cp}
\end{eqnarray}

and along with (\ref{Gamma}) we can identify 
\begin{eqnarray}
v&=&-{\frac{\partial\cm}{\partial\cp}}\Big/{\frac{\partial\cm}{\partial\cu}}=\left(\frac{\partial\cu}{\partial\cp}\right)_\cm\label{vGamm1},\\
\Gamma(v)&=&\frac{\partial \cm}{\partial\cu}\label{vGamm2}.
\end{eqnarray}

The RHS of these equations are homogeneous functions of degree 0 with respect to $\cu$ and $\cp$, and then depend only on the combination $\cp/\cu$. Consequently, from equation (\ref{vGamm1}) we can find $\cp/\cu$ as function of $v$, and from (\ref{vGamm2}) we can find explicitly the form of the function $\Gamma(v)$. Finally, taking into account that $v=\eta/\beta$ and $\mu=\beta/\Gamma$ we obtain the invariant $\mu$ as a function of $(\beta,\eta)$.

\section{Some Examples}

In this section we will apply our previous discussion and review some important cases: the relativistic dispersion relation and a relevant example found in the phenomenological discussions beyond that frame.

The relativistic case has been already discussed in detail in the previous sections, but we will analyze it briefly in terms of the arguments developed in the previous one. Let us begin from the relativistic dispersion relation $\cu^2=m^2+\cp^2$. It is clear that the function $\cm=\sqrt{\cu^2-\cp^2}$ corresponds to equation (\ref{mgen}) satisfying (\ref{clairautm}). Following the steps at the end of the previous section, using equations (\ref{vGamm1}) and (\ref{vGamm2}) we obtain that 
\begin{eqnarray}
v&=&\frac{\cp}{\cu},\\
\Gamma &=&\frac{1}{\sqrt{1-(\cp/\cu)^2}}=\gamma.
\end{eqnarray}

Finally, making $v=\eta/\beta$ we find that

\begin{equation}
\mu=\frac{\beta}{\Gamma}=\sqrt{\beta^2-\eta^2},
\end{equation}

as expected. Next, we discuss a more interesting case that is used as a probe of dispersion relation beyond the Lorentz Symmetry\cite{Mattingly, Piran, Ellis, Magueijo, Garay}, and that can be considered as a low energy limit of some models \cite{Smolin, Amelino-Smolin, Alfaro, Pullin}. A typical way to write this relation is

\begin{equation}\label{MDR}
E^2=m^2+p^2+\alpha l_p E^3,
\end{equation}  

where $\alpha$ is a dimensionless constant and $l_p$ a constant with length dimensions typically of order of the inverse of Planck scale $E_p=10^{19}\text{ GeV}$, such that for energies $E$ far below $E_p$ the above relation is accurately approximated by the relativistic one. 

As there is still no consensus on whether black holes should be considered fundamental objects or composite systems, we are going to consider how our arguments apply in each case. If one considers the black hole as a fundamental object, then the dispersion relation (\ref{MDR}) is applied directly, namely in our notation

\begin{equation}
\cu^2=\cm^2+\cp^2+\alpha l_p\cu^3.
\end{equation}

However, this dispersion relation clearly does not satisfy the condition imposed by (\ref{clairautm}), and then our argument discards this dispersion relation directly, as the homogeneity property is not fulfilled. This result is consistent with the fact that if we consider the black hole as a fundamental object, then clearly the LIV term in (\ref{MDR}) can be large, as its mass can widely exceed $M_p$, i.e. the Soccer Ball Problem would be present \cite{Soccerball2002}.

On the other hand, if we consider the black hole as a composite object, we can find the function $\cm$ of equation (\ref{Mgen}) compatible with the given dispersion relation and satisfying the homogeneity property (\ref{clairautm}) using the following implicit form

\begin{equation}
(1-\delta)\cm^2=\cu^2-\cp^2-\delta\frac{\cu^3}{\cm},
\end{equation}

where $\delta$ is a dimensionless constant that cannot depend upon $\cm$. Notice that we have chosen the coefficients so that $\cu=\cm(\cu,0)$, as required. If we consider this dispersion relation as a deviation from the relativistic one, then  $\delta\equiv\alpha l_p m<<1$. Calculating the invariant $\mu$ in the limit $\delta\to0$, which constitutes a small departure  from the relativistic scenario, we find at first order in $\delta$ that

\begin{equation}
v=\frac{\cp}{\cu}\left(1+\frac{3\delta}{2}\frac{1}{\sqrt{1-(\cp/\cu)^2}}+...\right),
\end{equation}

or conversely,

\begin{equation}
\frac{\cp}{\cu}=v\left(1-\frac{3\delta}{2}\gamma +...\right).
\end{equation}

On the other hand we obtain

\begin{equation}
\Gamma(v)=\gamma\left[1+\frac{\delta}{2}(1-3\gamma+2\gamma^3)\right]+...
\end{equation}

and then the function $\mu(\beta,\eta)$ in this case is given by

\begin{equation}
\mu=\sqrt{\beta^2-\eta^2}-\frac{\delta}{2}\left(\sqrt{\beta^2-\eta^2}-3\beta+2\frac{\beta^3}{\beta^2-\eta^2}\right).
\end{equation}

Notice that, in accordance with the previous discussion, we indeed have that $\mu(\beta,0)=\beta$ also in this case. This invariant $\mu$ must be a direct consequence of the symmetry that would replace the Lorentz Transformations, and one should be able to identify the new transformations.

In this example $m$ is an arbitrary scale that can be fixed by convenience. If we choose $m$ to be of the same order as the mass of the system $\cm$ then it is clear that, in order to maintain the phenomenological consistency for black holes with macroscopic masses, that is, in order to avoid the Soccer Ball Problem, $\alpha<<1$, and then (\ref{MDR}) implies that the LIV scale should be far below the Planck scale $l_p$. Namely, the LIV term appearing in (\ref{MDR}) needs to be strongly suppressed by a scale far above $E_p$.  This conclusion is in agreement with the results in \cite{Relative-Soccer}, where the Soccer Ball Problem is investigated in the context of Relative Locality, and it is shown that, for a composite system, the first correction to the relativistic dispersion relation is suppressed not by $E_p$ but by $NE_p$, being $N$ the number constituents. Thus, the present analysis shows that the modified dispersion relation \ref{MDR} is consistent only if one indeed considers the black hole as a composite system, and the LIV term is suppressed by a scale above $E_p$, avoiding the Soccer Ball Problem. Moreover, one could think, for instance, on the existence of additional adiabatically invariant quantities defining the system, namely, entropy could depend not only on $\cu$ and $\cp$ but also on another extensive quantity \cite{Ghosh}. In the framework of black holes, these could be known quantities as angular momentum or charge, or quantities characterizing the microscopic state of the system.

\section{Final Remarks and Outlook}

In this work we have studied the thermodynamics of a moving Schwarzschild black hole identifying the temperature and entropy in a relativistic scenario. Furthermore, we have set arguments in a framework relating invariant geometrical quantities under global spacetime transformations and the dispersion relation of the system. Next, we have extended these arguments in order to consider more general dispersion relations. The formalism just shown may be interesting for analyzing LIV phenomenology in massive objects presumably structureless as black holes. As we discussed at the end of section IV, there are dispersion relations used in discussions on LIV phenomenology that present some difficulties in virtue of our analysis. A possible solution to such drawbacks could be to consider more general black holes or presume certain internal structure yet unknown. We expect to exploit this last idea in next works, firstly by considering rotating and charged black holes, and afterwards by considering new degrees of freedom that may match our arguments with phenomenological observations in a natural way.

\begin{acknowledgments}
The authors would like to acknowledge the inspiring comments of Aurelio Grillo and Fernando Mendez, and the support of FONDECYT Chile, through grant 1140243.

\end{acknowledgments}

%
%
%
%
%
%
%
%
%
%
%
%
%
%
%
%
%
%
%
%
%
%

\bibliographystyle{apsrev4-1}

\end{document}